\documentclass[doublecol]{epl2} 

\usepackage{graphics}
\usepackage{epsfig}
\usepackage{amssymb}
\usepackage{subfigure}
\usepackage{pifont}

\title{Athermal analogue of sheared dense Brownian suspensions}
\shorttitle{Athermal analogue of sheared dense Brownian suspensions} 

\author{Martin Trulsson \and Mehdi Bouzid \and Jorge Kurchan \and Eric Cl\'ement \and Philippe Claudin\\ \and Bruno Andreotti}
\shortauthor{M. Trulsson \etal}
\institute{Physique et M\'ecanique des Milieux H\'et\'erog\`enes, UMR 7636 ESPCI -- CNRS -- Univ.~Paris-Diderot -- Univ.~P.M.~Curie, 10 rue Vauquelin, 75005 Paris, France.}

\pacs{83.80.Hj}{Rheology of suspensions}
\pacs{47.57.Gc}{Granular flows -- Complex fluids}
\pacs{47.57.Qk}{Rheology of complex fluids}
\pacs{82.70.Kj}{Suspensions}

\abstract{
The rheology of dense Brownian suspensions of hard spheres is investigated numerically beyond the low shear rate Newtonian regime. We analyze an athermal analogue of these suspensions, with an effective logarithmic repulsive potential representing the vibrational entropic forces. We show that both systems present the same rheology without adjustable parameters. Moreover, all rheological responses display similar Herschel-Bulkley relations once the shear stress and the shear rate are respectively rescaled by a characteristic stress scale and by a microscopic reorganization time-scale, both related to the normal confining pressure. This pressure-controlled approach, originally developed for granular flows, reveals a striking physical analogy between the colloidal glass transition and granular jamming.
}

\begin{document}

\maketitle

\begin{figure}[ht!]
\centerline{\includegraphics{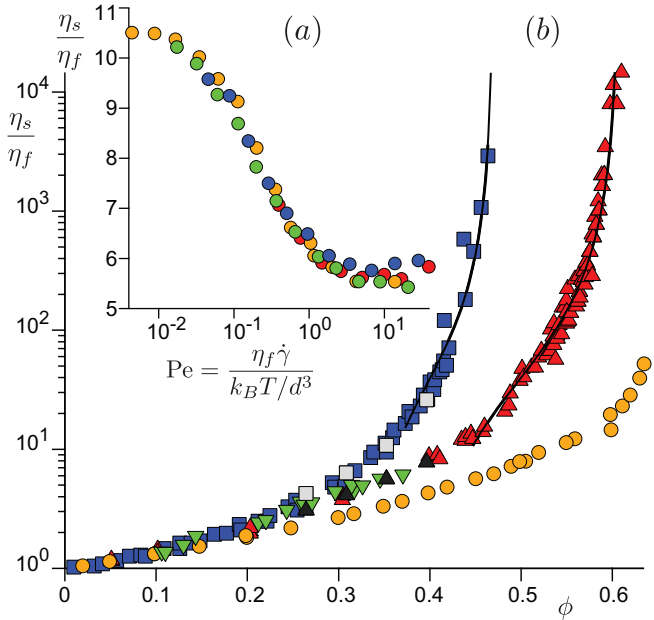}}
\vspace{-5 mm}
\caption{
{
(a) Relationship between the rescaled suspension viscosity $\eta_s/\eta_f$ and the P\'eclet number $\rm Pe$ for sterically stabilized poly(methylmethacrylate) particles of different sizes: $85\,{\rm nm}$ (orange $\circ$), $141\,{\rm nm}$ (green $\circ$), $204\,{\rm nm}$ (blue $\circ$) and $310\,{\rm nm}$ (red $\circ$) dispersed in silicon oil -- data from \cite{CK86}. From the asymptotic value of the viscosity at large $\rm Pe$, one extracts the effective volume fraction $\phi \simeq 0.31$ following the law calibrated in the non-Brownian limit (red $\vartriangle$). (b) Rescaled viscosity $\eta_s/\eta_f$ of non-brownian and brownian suspensions as a function of $\phi$: data from experiments and previous numerical models. Red $\vartriangle$: compilation of data for isodense non-brownian suspensions \cite{OBR06,BDEL10}. Green $\triangledown$: direct numerical simulations of non-Brownian suspensions \cite{PBMB13}. Orange $\circ$: compilation of numerical data obtained by Stokesian dynamics \cite{SB2001}. Blue $\square$: compilation of data for colloidal suspensions \cite{HW95}. For all these experimental data, $\phi$ has been re-determined using the large P\'eclet viscosity, as described in (a). Black $\vartriangle$ and grey $\square$ respectively show the asymptotic values of the viscosity at small and large $\rm Pe$ determined from the data in \cite{CK86}. The solid lines show the best fit of the data by the functional form obtained numerically (see Fig.~\ref{sfig2}).
}
}
\vspace{-4 mm}
\label{fig0}
\end{figure}

Amorphous materials \cite{K05} such as granular packings~\cite{AFP13}, colloidal suspensions~\cite{PvM85,MW95,HW95,MW12,CBML03} or glassy molecular systems \cite{EAN96,DS01,MKK09} display a severe increase in their viscosity before solidifying. With glass-forming liquids, an amorphous solid is obtained by rapidly cooling below the freezing temperature. In the case of systems of hard particles with size dispersion, an emergent solid-like behavior as the packing fraction $\phi$ is increased and the amorphous character stems from geometrical disorder. In spite of essential differences in microscopic interactions, the existence of a universal scenario leading to a dynamical slowing down and to the emergence of rigidity~\cite{OT07} remains a vivid issue~\cite{LN98,MKK09,IBS13,BSPBST14,BSSPBST15}. Colloids, by the nature of their interparticle interactions and their size, are at the cross-roads between molecular thermal systems and athermal granular materials~ \cite{IBS12,W76,KD59,OBR06,BDEL10,BGO11,ABH12,LDW12, Midi04,JFP06}. In the dense regime, colloidal suspensions bear all the phenomenological complexity shared by glass-forming materials. The strong dynamical slowing down is associated with particle trapping in cages formed by their neighbours \cite{MW12,FC09}. Because of this relatively simple picture, colloids are often considered as paradigmatic systems to approach the general issues of dynamical arrest, glass transition and non-Newtonian rheology observed in complex fluids. 

\begin{figure}[t!]
\includegraphics{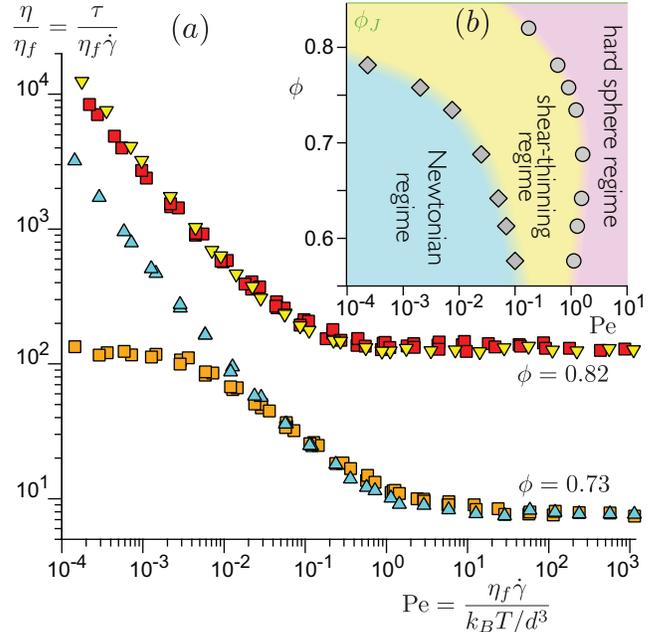}
\vspace{-5 mm}
\caption{(a) Rheological behaviour of a sheared Brownian suspension (squares) and its non-Brownian analogue (triangles): relationship between the particle borne viscosity $\eta$, rescaled by $\eta_f$ and the P\'eclet number $\rm Pe$, for two volume fractions $\phi$. Symbols: Brownian particles for $\phi=0.73$ (orange $\square$) and $\phi=0.82$ (red $\square$); Non-Brownian particles coated with a soft layer, whose elasticity is described by the logarithmic potential, for $\phi=0.73$ (turquoise$\vartriangle$) and $\phi=0.82$ (yellow $\triangledown$). (b) Rheological regimes in the plane $\rm Pe$ \emph{vs} $\phi$. Newtonian regimes at small and large $\rm Pe$ correspond to plateaus in panel (a),
{whose respective viscosities is denoted by $\eta_B(\phi)$ and $\eta_H(\phi)$.}
}
\vspace{-4 mm}
\label{fig1}
\end{figure}

The overall viscosity of an athermal suspension results from two contributions: the fluid and the particles. It is known since the original Einstein's 1906 article \cite{Einstein1906} and subsequent theories \cite{Batchelor70} and simulations \cite{FB00} using Stokesian dynamics (Fig.~\ref{fig0}), that there is an increase with particle density in the system's viscosity  due to an increase in hydrodynamic interactions. However, this effect does not account for  the drastic dynamical slowing-down measured experimentally \cite{KD59,OBR06,BDEL10,BGO11}: even for the state of the art numerical simulations \cite{SB2001,BLEW14}, the viscosity is strongly underestimated in the dense limit (Fig.~\ref{fig0}). As shown in \cite{BGO11}, a stress related to collective particle effects and  in the complexity of the particle trajectories under geometrical constraints, is actually responsible for the diverging viscosity \cite{ABH12,LDW12}.

In this letter, we extend this line of thought to dense Brownian suspensions using a model that focusses on collective particle phenomena, the physics at play becoming close to that of the glass transition \cite{IBS12,IBS13}.
We go beyond direct numerical simulations \cite{PBMB13}, which are very accurate, but limited in practice to the semi-dilute regime i.e. to suspension viscosities smaller than, say, ten times that of the fluid {(Fig.~\ref{fig0})}. We show that our model allows us to bridge the gap between the rheology of Brownian and athermal dense suspensions. To this end, we use an exact expression previously derived for quiescent thermal hard-spheres close to the jamming transition~\cite{BW06} in which an effective potential accounts for vibrational entropic forces, i.e.  a  non-thermal analogue for Brownian suspensions. We extend this concept for rheology, and show that these systems share similar non-Newtonian behavior under shear once energy and time are properly rescaled.

\section{Numerical method}
The rheology of suspensions is investigated by means of numerical simulations based on a coupling between particles, described using a discrete element method, and the surrounding fluid, described at the continuous level as in \cite{TAC12}.
{
We consider a two-dimensional system, constituted of  $N \simeq 10^3$ circular particles labeled by an index $i$, with diameters $d_i$ randomly chosen in a flat distribution between $0.5\;d$ and $1.5\;d$, where $d$ is the mean particle size. Such a polydispersity is a compromise ensuring that the sample remains statistically homogeneous and does not crystalize when sheared. The particles are frictionless and confined in a plane shear cell composed of two rough walls moving along the $x$-direction at opposite and constant velocities, leading to a homogeneous shear rate in the bulk denoted as $\dot \gamma$. The walls are made of similar particles, glued together. The distance between the walls in the perpendicular  $y$-direction is fixed to ensure a constant volume fraction $\phi$. Periodic boundary conditions are used along the $x$-direction. 
}

{
The particles are immersed in a fluid of viscosity $\eta_f$ and we describe its flow in the mean field by an imposed affine velocity field $\dot \gamma y \, {\bf e}_x$. where ${\bf e}_x$ is the unit vector along the $x$-axis. This simplified description of hydrodynamics is inaccurate in the dilute regime, as hydrodynamics interactions between particles then become dominant \cite{FB00}, but sufficient in the regime where the particle-borne stress becomes larger than the fluid-borne stress \cite{ABH12}. This is directly proved by Fig.~(\ref{fig0}), which compares predictions ignoring the particle-borne stress (orange circles) to measurements (red triangles). The contact between two particles is modeled as a harmonic spring, whose spring constant $k_n$ is large enough to obtain results independent of $k_n$ (the rigid limit). In practice, simulations are performed with $k_n\simeq 10^4\,P$, where $P$ is the average pressure associated with these contacts (see below). Importantly, $P$ is homogeneous all through the system and constant over time -- although fluctuating as a dynamical quantity -- in the steady regime we are interested in.}

{
We assume that the evolution of the particle $i$ is governed by the overdamped Langevin equation:
\begin{equation}
- \frac{3\pi\eta_f}{(1-\phi)}\boldsymbol{\delta u}_i + \sum_{j\neq i} \boldsymbol{f}_{ij} + \boldsymbol{\xi}_i =  0,
\label{langevin}
\end{equation}
from which the particle velocity $\boldsymbol{u}_i$ and position $\boldsymbol{r}_i$ are computed at each time step. In this expression, the first term corresponds to the drag force, due to the interaction of the particle with the surrounding fluid. It is proportional to the non-affine particle velocity $\boldsymbol{\delta u}_i \equiv  \boldsymbol{u}_i - \dot \gamma y {\bf e}_x$ and leads to a dissipation directly related to the suspension viscosity and that is quadratic in $\boldsymbol{\delta u}$ \cite{ABH12}. The second term involves the contact forces $\boldsymbol{f}_{ij}$ felt by particle $i$ from its neighbours $j$. We introduce $r_{ij} = |\boldsymbol{r}_j - \boldsymbol{r}_i|$ the distance between the particles $i$ and $j$, $\boldsymbol{n}_{ij}$ the normal vector pointing from $i$ to $j$, and $d_{ij} = (d_j+d_i)/2$ their average diameter. With these notations, we write the contact force as the derivative of an interaction potential $\boldsymbol{f}_{ij}= - \mathcal{U}'(r_{ij}) \boldsymbol{n}_{ij}$, where here the potential is harmonic: $\mathcal{U}(r_{ij}) = \frac{1}{2} k_n (r_{ij} - d_{ij})^2$ when $r_{ij} < d_{ij}$, and $\mathcal{U}(r_{ij}) = 0$ when the particles do not touch ($r_{ij} > d_{ij}$). Simulations of the Brownian system are performed submitting the particles to random forces $\boldsymbol{\xi}_i$. We take for them a white noise of variance $\langle\boldsymbol{\xi}_i(t)\boldsymbol{\xi}_j(t')\rangle=6\pi\eta_f k_BT/(1-\phi)\delta_{ij}\delta(t-t')$, which represents thermal fluctuations of the surrounding liquid --~$T$ is the temperature and $k_B$ the Boltzmann constant. Technically, the numerical time step is chosen to be a small fraction (typically $\simeq 1/20$) of the microscopic timescale $3\pi\eta_f/[(1-\phi)k_n]$. Tests have been performed to ensure that our results are independent of this value, as long as it is small enough. Finally, the stress components are measured as $\sigma_{\alpha\beta}=\frac{1}{2 A} \sum_{ij} {f}_{ij}^{\alpha} {r}_{ij}^{\beta}$ over an area $A$ in the center of the cell, large enough to get a result independent of $A$. The shear stress is $\tau=\sigma_{xy}$ and the confining pressure is given by $P=\sigma_{yy}$. It is worth noting that normal stress difference is found to be smaller than the $1\%$ error bars.}

\begin{figure}[t!]
\centerline{\includegraphics{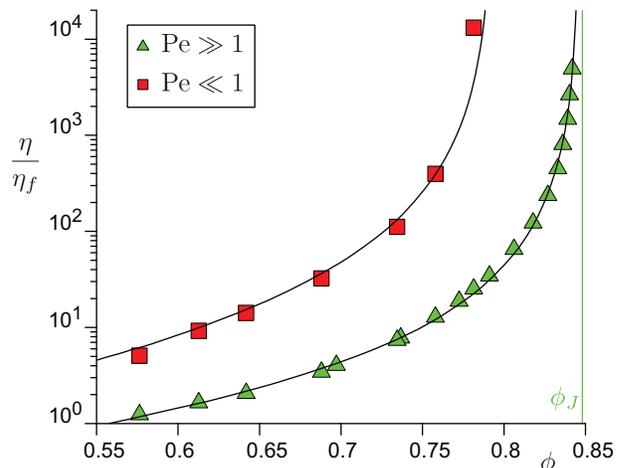}}
\vspace{-5 mm}
\caption{{Rescaled viscosity corresponding to the Newtonian plateaus at small (red $\square$) and large (green $\vartriangle$) P\'eclet regimes in Fig.~\ref{fig1}a, as a function of the particle volume fraction. Black line on green triangles: phenomenological fit by a law diverging as $\eta_H(\phi) \propto \eta_f (\phi_J-\phi)^{-2}$. Black line on red squares: deduced from that on the green triangles by Eq.~\ref{etaB}. }}
\label{sfig2}
\vspace{-4 mm}
\end{figure}

\begin{figure}[t!]
\includegraphics{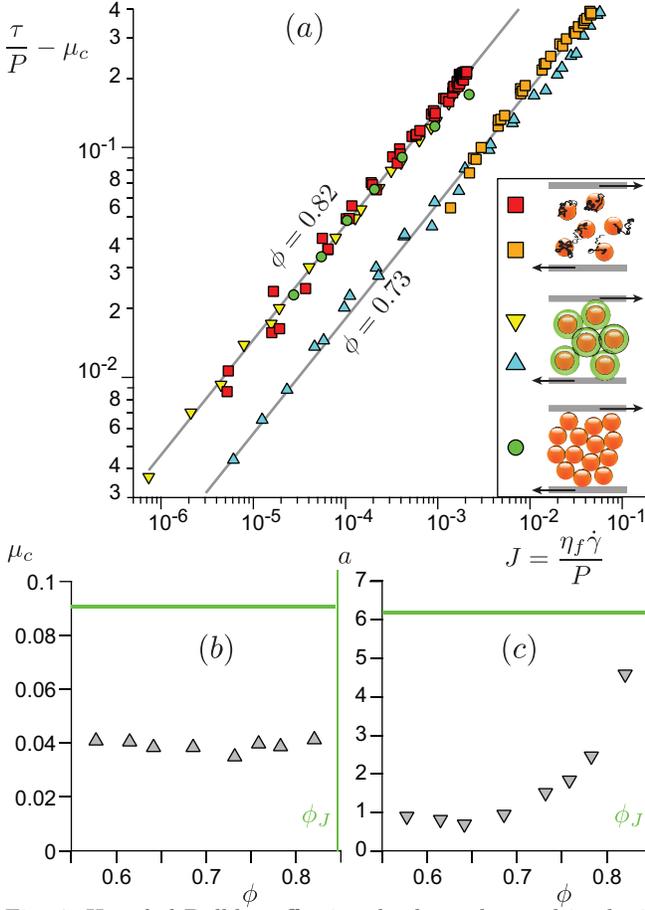}
\vspace{-5 mm}
\caption{Herschel-Bulkley effective rheology observed in the intermediate $\rm Pe$ regime (shear thinning region shown in Fig~\ref{fig1}). (a) Reduced effective friction coefficient as a function of the dimensionless viscous number. Same symbols as in Fig.~\ref{fig1}. Green circles are reference data obtained for suspensions of non-Brownian rigid particles. Note that, in this case, each point corresponds to a different value of $\phi$. Grey lines: fit of Eq.~\ref{HB} with a slope $\nu=1/2$.
{
Critical friction coefficient $\mu_c$ (b) and multiplicative factor $a$ (c) as a function of $\phi$ for the athermal analogous system (triangles) and for the reference non-Brownian rigid particles (green line).
}
}
\vspace{-4 mm}
\label{fig3}
\end{figure}

When slowly sheared, Brownian particles preserve their neighbours over long times and one may assume a time-scale separation between the rapid (near-contact) collisions and the slow, collective evolution of the particle spatial configurations. Close enough to  $\phi_J$, the phase-space volume can be expressed as the product of the gaps between the particles, implying that for hard spheres, the interactions due to collisions \cite{FC06,AL03} yield, for an inter-particle distance $r$, an effective potential of the form~\cite{BW06}.
\begin{equation}
\varphi(r) = k_BT \ln \left( \frac{r-d}{d} \right) .
\label{varphilog}
\end{equation}
The interaction forces are then inversely proportional to the gap $\epsilon = r-d$ between the particles, as can be easily understood considering that particles exchange momentum in their frequent collisions with velocities proportional to $\sqrt{T}$.

{
The validity of the effective logarithmic potential (\ref{varphilog}) was assessed for an unsheared system, in the glassy regime, near random close packing \cite{BW06}. In this paper, we test whether this approach can be extended to rheology by comparing simulations of Brownian rigid spheres and non-Brownian spheres coated with a soft layer whose elasticity is described by $\varphi(r)$. Formally, expression (\ref{varphilog}) is only valid for true hard sphere interactions between particles. For particles with a large but finite elasticity deriving from a potential $\mathcal{U}$, the above expression is generalized as:
\begin{equation}
\varphi(r_{ij}) = k_B T \ln \left[ \frac{\int_0^{r_{ij}} e^{-\frac{\mathcal{U}(s)}{k_B T}}\, {\rm d}s}{\int_0^{r_{ij}} \,{\rm d}s}  \right] .
\label{varphiloggeneralised}
\nonumber
\end{equation}
Analogous athermal simulations are performed by replacing the sum of contact and random forces by the deterministic effective force $- \sum_{j\neq i}\varphi' (r_{ij}) \boldsymbol{n}_{ij}$ in (\ref{langevin}). As standardly done in the numerical handling of long range interactions, a cut-off distance $\epsilon_c$ is introduced: the inter-particle force is given by $\varphi'(r_{ij})$ for $r_{ij}<d_{ij}+\epsilon_c$ and vanishes for $r_{ij}>d_{ij}+\epsilon_c$. Except in the last section, the following results are independent of $\epsilon_c$, provided $\epsilon_c$ and $\phi$ are large enough ($\epsilon_c > d/10$ and $\phi > 0.5$ in practice).
}


%
%

The systems are controlled by two dimensionless numbers: the particle volume fraction $\phi$ and the P\'eclet number  ${\rm Pe} \equiv \eta_f \dot \gamma /(k_B T d^{-3})$, which compares viscous dissipation and thermal fluctuations at the scale $d$ of the particle. The particle-borne viscosity $\eta$ is deduced from the shear stress $\tau$ and originates from the contact forces between particles. It is important to notice that it does not include the fluid-borne viscosity.
{To emphasize this point, the viscosity of the whole suspension is denoted as $\eta_s$, as in Fig.~\ref{fig0}, which shows experimental data and previous numerical models.}

\begin{figure}[t!]
\includegraphics{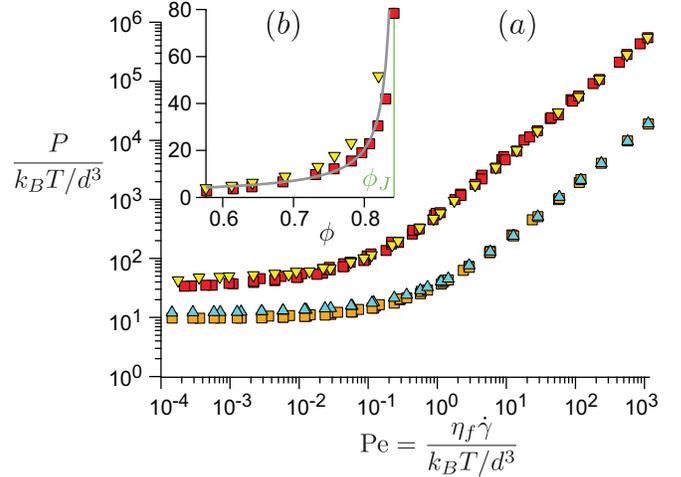}
\vspace{-5 mm}
\caption{Rescaled pressure as a function of $\rm Pe$. Same symbols as in Fig.~\ref{fig1}. Inset: zero-shear pressure as a function of $\phi$ for the Brownian suspension (squares) and its non-Brownian analogue (triangles). Line: phenomenological fit with a divergent behaviour in $(\phi_J-\phi)^{-2}$ close to jamming.}
\vspace{-4 mm}
\label{fig2}
\end{figure}

\section{Strongly sheared regime}
As observed experimentally \cite{Krieger1972, IBS12,MW12}, the simulated Brownian suspension presents three different regimes when varying the P\'eclet number at constant  $\phi$ (Fig.~\ref{fig1}b): a Newtonian, quasi-equilibrium regime at low  $\rm Pe$, a shear-thinning regime at intermediate  $\rm Pe$ and a second Newtonian regime at $\rm Pe>1$. As shown in Figs.~\ref{fig1}a and \ref{fig2}, the non-Brownian analogue is in quantitative agreement for the behavior of rescaled shear stress and pressure as functions of $\rm Pe$, throughout the intermediate, shear-thinning regime, and at large $\rm Pe$. This collapse of rheological curves is obtained without adjustable parameter whatever the value of $\phi$, above some cross-over $\rm Pe$ represented in Fig.~\ref{fig1}b.

At large P\'eclet number (pink zone in Fig.~\ref{fig1}b), thermal agitation has a negligible influence compared to the viscous drag and contact forces, and  the Brownian suspension asymptotically behaves as a suspension of non-Brownian rigid spheres. Similarly, the potential $\varphi(r)$ as a negligible effect on the non-Brownian analogue, which therefore behaves as the very same suspension non-Brownian rigid spheres. The viscosity  diverges at a critical jamming concentration $\phi_J$, defined as the maximal volume fraction reachable under a permanent quasi-static shear without allowing overlap between particles. As shown in Fig.~\ref{sfig2}, when approaching this jamming point, the viscosity $\eta_H \sim   (\phi_J-\phi)^{-2\alpha}$, with  $2\alpha \sim 2 \pm 0.5$ \cite{OBR06,BGO11}. This is consistent with a dynamics that is, in this limit, entirely controlled by steric effects associated with the non-affine collective motions: a particle statistically travels a path of mean length $d(\phi_J-\phi)^{-\alpha}$ in order to move a distance  $d$ as the crow flies \cite{ABH12,LDW12}.
{These cooperative effects are purely geometrical and do not involve interactions between elasto-plastic events as observed in soft elastic systems \cite{BSPBST14,BSSPBST15}.}

\begin{figure*}
\includegraphics{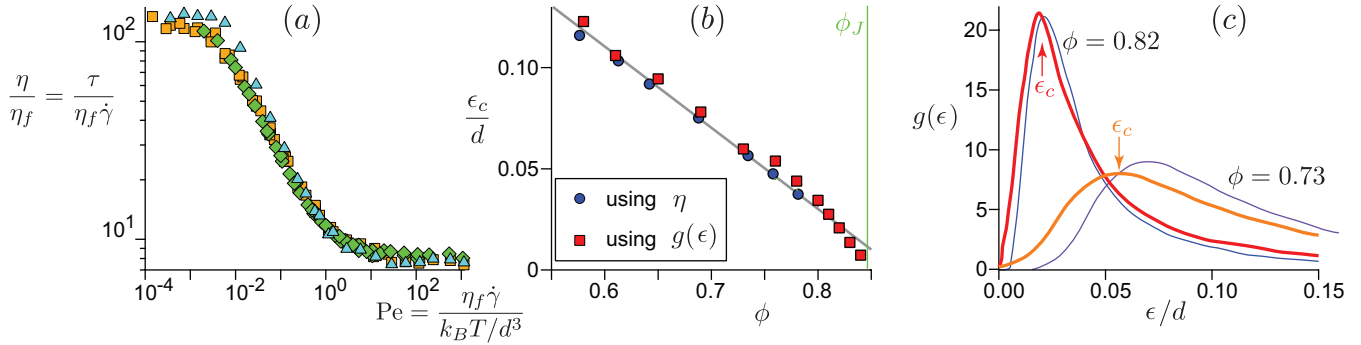}
\vspace{-5 mm}
\caption{
{
(a) Rescaled viscosities as a function of $\rm Pe$ for numerical simulations of various system sizes at $\phi=0.73$. Orange $\square$:  system of 800 Brownian particles. Green $\diamond$:  system of 1800 Brownian particles. Blue $\vartriangle$: analogous system of 800 non-Brownian particles whose interactions derive from the potential $\varphi$ and are cut at $\epsilon_c=0.055\;d$ (soft layer coating).
}
(b) Cut-off distance $\epsilon_c$, in particle size unit, limiting the range of the logarithmic interaction potential, \emph{versus} $\phi$. Blue $\circ$: value of $\epsilon_c$ computed by matching the viscosities in the Newtonian regime at small $\rm Pe$. Red $\square$: value of $\epsilon_c$ deduced from the maximum of the pair-correlation function, see next panel. Black line: linear fit on the blue points.
(c) Position averaged pair correlation function $g$ as a function of the gap $\epsilon$. Lines: Brownian suspensions at $\phi=0.73$ (orange) and $\phi=0.82$ (red); non-Brownian analogue at $\phi=0.73$ (violet) and $\phi=0.82$ (blue). $\epsilon_c$ is the value of the gap at the first peak in the correlation.}
\vspace{-4 mm}
\label{fig4}
\end{figure*}

In the intermediate range of $\rm Pe$ (yellow zone in Fig.~\ref{fig1}b), the shearing timescale  $\dot \gamma^{-1}$  is small compared to the thermal equilibration time. The imposed shear drives the system out of equilibrium but thermal effects are still important. In this regime, pressure is therefore dominated by thermal collisions. This suggests to analyze the data in a way similar to the pressure-controlled approach developed for dry granular flows \cite{Midi04,JFP06}, and generalized for dense granular suspensions \cite{BGO11,TAC12}. If, for a given volume fraction, the value of $P$ is the relevant characteristic of the energy landscape, the shear stress $\tau$ should be rescaled to form a dimensionless yield parameter $\tau/P$. As the dynamics is overdamped, the relevant time-scale for plastic rearrangements is simply $\eta_f/P$ and the relevant dimensionless shear rate is the viscous number $J \equiv \eta_f \dot \gamma/P$. In the reference case of suspensions of non-Brownian rigid particles, $\tau/P$ and $\phi$ are functions of $J$, and the rheology follows a Herschel-Bulkley relation of the form: 
\begin{equation}
\frac{\tau}{P}= \mu_c+a J^\nu,
\label{HB}
\end{equation}
where $\nu$, $\mu_c$ and $a$ are constitutive parameters of the system \cite{BGO11,TAC12}. These reference data are displayed in Fig.~\ref{fig3}a in a plot of the reduced friction $\tau/P - \mu_c$ \emph{vs} $J$, in order to provide evidence of a scaling law behaviour.
{
The data from the simulations of the Brownian suspensions and their athermal analogue are also plotted in the same graph, for two values of the volume fraction.  The critical friction coefficient $\mu_c$ has been extracted from the fit of the athermal analogous data, for which this intermediate regime is not interrupted at small P\'eclet (see below), allowing for a precise adjustment. Both Brownian and athermal analogous data collapse when subtracting this very same $\mu_c$.
}
For all these cases, the phenomenological law (\ref{HB}) is well verified over 4 decades in viscous number, with a convincingly constant exponent  $\nu \simeq 0.5$. Interestingly, as $\phi$ approaches $\phi_J$, all the reduced data apparently converge to the reference line. Accordingly, the multiplicative factor $a$ tends to the reference value at $\phi_J$ (Fig.~\ref{fig3}c). These results support the idea that the confinement pressure and the steric effects play the same role in all systems, and do control the statistical properties of the particles trajectories and consequently the dissipation. $\mu_c$ is found to be independent of $\phi$ (Fig.~\ref{fig3}b).
{
However,  in the case of  Brownian suspensions, the critical friction values ($\mu_c \simeq 0.04$) differs from the value obtained for the reference model of non-Brownian suspensions of rigid particles ($\mu_c \simeq 0.09$).
}

\section{Newtonian regime}
Close to thermal equilibrium, at small P\'eclet, shear is slow enough to allow for thermally activated changes of particle configuration. In this low $\rm Pe$ regime, a Newtonian plateau is observed, corresponding to a well defined viscosity $\eta_B$ of the Brownian suspension (Fig.~\ref{fig1}a). As shown in Fig.~\ref{fig4}a, the viscosity rapidly increases with $\phi$. This behavior resembles that of the viscosity associated with the large $\rm Pe$ regime, $\eta_H$, which diverges as $\phi \to \phi_J$, but its origin is different: here we are probing the actual glass transition -- probably only a crossover, at least in this two-dimensional case -- which takes place at $\phi<\phi_J$ for sufficiently low  shear rates. The non-Brownian analogue developed here using the interaction potential (\ref{varphilog}) is able to reproduce the behavior of pressure at low $\rm Pe$ (Fig.~\ref{fig2}) but not the Newtonian viscosity (Fig.~\ref{fig1}a). However, the analogy between Brownian and non-Brownian systems can be extended treating   $\epsilon_c$  as an {\em ad hoc}  parameter, allowing one to capture in a phenomenological way, some elements of the glass transition. With a small value of $\epsilon_c$, the non-Brownian analogue is now constituted of particles coated a finite thickness soft layer and presents a Newtonian behaviour below the Herschel-Bulckley regime (Fig.~\ref{fig4}a): it behaves as a non-Brownian suspension of hard particles of diameter $d + \epsilon_c$, whose effective volume fraction is $\left( 1 + \epsilon_c/d \right)^2 \phi$. The viscosity of the Brownian suspension therefore matches that of the non-Brownian suspension if the relation:
\begin{equation}
\eta_B(\phi)=\eta_H \left [ \left(1+\frac{\epsilon_c}{d} \right)^2 \phi \right]
\label{etaB}
\end{equation}
is obeyed. A second independent route to obtain $\epsilon_c(\phi)$ is to determine the effective pair correlation function $g$ from equilibrium (un-sheared) simulations of the Brownian system. The particle position is averaged over the cage persistence time (chosen to correspond to the inflection point in the single particle diffusion curve $\left< r^2 \right >$ \emph{vs} $t$). This function presents a sharp peak at short gaps that provides a second estimate for $\epsilon_c$ (Fig.~\ref{fig4}c). Fig.~\ref{fig4}b shows that the two definitions of $\epsilon_c$ match quantitatively. Importantly, the pressure of the non-Brownian analogue only captures the dynamic pressure associated to shearing, but not the thermal pressure, which must be added.

The collapse on an Herschel-Bulkley rheology for the dense suspensions in the intermediate P\'eclet regime leads to a striking remark: the system always `feels' the presence of a yield stress even though, at low $\rm Pe$, the rheology turns Newtonian --~at least for the less dense suspensions. The value $\mu_c$ of the dynamical friction extrapolated at vanishing shear rate is found independent of the packing fraction. This points to the existence of microscopic mechanisms leading to plastic flow, common to all dense Brownian suspensions and by the sake of our analogy, also common to soft-systems~\cite{LC09, BCA09,LP09}. The fact that the threshold value is significantly different (by a factor of $\simeq 1/2$) from the critical dynamical friction obtained for hard-sphere suspensions indicates furthermore, that hard-sphere and soft-sphere systems are essentially different as far as microscopic flow processes are concerned \cite{Bouzidpreprint2014}. Finally, the  common behavior of Brownian and non-Brownian hard sphere systems at large shear rates may offer a practical way to measure accurately the volume fraction of suspensions (Fig.~\ref{fig0}a). This measurement is always an experimental difficulty, and one could cleverly make use of the well calibrated hard-sphere regime.

BA is supported by Institut Universitaire de France. This work is funded by the ANR JamVibe and a CNES research grant.

\vspace{-4 mm}

\end{document}